# Electronic Properties of Graphene Encapsulated with Different Two-Dimensional Atomic Crystals

A. V. Kretinin,*,[1] Y. Cao,[1] J. S. Tu,[1] G. L. Yu,[2] R. Jalil,[1] K. S. Novoselov,[2] S. J. Haigh,[3] A. Gholinia,[3] A. Mishchenko,[2] M. Lozada,[2] T. Georgiou,[2] C. R. Woods,[2] F. Withers,[1] P. Blake,[1] G. Eda,[4] A. Wirsig,[5] C. Hucho,[5] K. Watanabe,[6] T. Taniguchi,[6] A. K. Geim[1,2] and R. V. Gorbachev[1]

[1]Centre for Mesoscience and Nanotechnology, University of Manchester, Manchester M13 9PL, UK

[2]School of Physics and Astronomy, University of Manchester, Oxford Road, Manchester, M13 9PL, UK

[3]School of Materials, University of Manchester, Oxford Road, Manchester, M13 9PL, UK

[4]Graphene Research Centre, National University of Singapore, 6 Science Drive 2, Singapore 117546

[5]Paul Drude Institut für Festkörperelektronik, Hausvogteiplatz 5-7, 10117 Berlin, Germany

[6]National Institute for Materials Science, 1-1 Namiki, Tsukuba, 305-0044 Japan

*Hexagonal boron nitride is the only substrate that has so far allowed graphene devices exhibiting micron-scale ballistic transport. Can other atomically flat crystals be used as substrates for making quality graphene heterostructures? Here we report on our search for alternative substrates. The devices fabricated by encapsulating graphene with molybdenum or tungsten disulphides and hBN are found to exhibit consistently high carrier mobilities of about 60,000 cm$^2$ V$^{-1}$ s$^{-1}$. In contrast, encapsulation with atomically flat layered oxides such as mica, bismuth strontium calcium copper oxide and vanadium pentoxide results in exceptionally low quality of graphene devices with mobilities of ~1,000 cm$^2$ V$^{-1}$s$^{-1}$. We attribute the difference mainly to self-cleansing that takes place at interfaces between graphene, hBN and transition metal dichalcogenides. Surface contamination assembles into large pockets allowing the rest of the interface to become atomically clean. The cleansing process does not occur for graphene on atomically flat oxide substrates.*

Until recently, the substrate of choice in microfabrication of graphene devices was oxidized Si wafers. This was due to their availability and versatility, excellent dielectric properties of thermally grown $SiO_2$, and easy visualization and identification of monolayer and bilayer graphene on top of such substrates.[1] However, it has soon become clear that the quality of graphene-on-$SiO_2$ devices was limited by several factors including surface roughness, adatoms acting as resonant scatterers and charges trapped at or near the graphene-$SiO_2$ interface.[1-3] Search for better substrates had started[4] and eventually led to the important finding that cleaved hexagonal boron nitride (hBN) provides an excellent substrate for graphene.[5, 6] Typically, graphene-on-hBN exhibits a tenfold increase in the carrier mobility, $\mu$, with respect to devices made on $SiO_2$.[5] This quality of graphene has made it possible to observe the fractional quantum Hall effect[6] and various ballistic transport phenomena.[7, 8] Although hBN is now widely used for making increasingly complex van der Waals heterostructures,[9-11] it remains unclear whether it is only the atomic flatness of hBN that is essential for electronic quality or other characteristics also play a critical role. Even more important is the question whether hBN is unique or there exist other substrates that may allow graphene of high electronic quality.

In this Letter we report on our studies of various layered materials as atomically flat substrates for making graphene devices and van der Waals heterostructures. By using transport and capacitance measurements, we assess the electronic quality of monolayer graphene encapsulated between transitional metal dichalcogenides (TMD), such as $MoS_2$ and $WS_2$, and several layered oxides such as muscovite mica, bismuth strontium calcium copper oxide (BSCCO) and vanadium pentoxide ($V_2O_5$), on one side and hBN on the other. As a reference for electronic quality, we use graphene-on-$SiO_2$ and hBN/graphene/hBN heterostructures. In the latter case, we can usually achieve $\mu$ of ~100,000 $cm^2 V^{-1} s^{-1}$ [7, 12] and, with using the 'dry-peel' transfer,[12] $\mu$ can go up to 500,000 $cm^2 V^{-1} s^{-1}$, allowing ballistic devices with scattering occurring mainly at sample boundaries.[7, 12] The $MoS_2$/graphene/hBN and $WS_2$/graphene/hBN structures are also found to exhibit high quality ($\mu$ ~60,000 $cm^2 V^{-1} s^{-1}$) and high charge homogeneity, which makes $MoS_2$ and $WS_2$ a good alternative to hBN. Regarding atomically flat oxides, their use results in dismal electronic quality, which is lower than that observed for atomically rough surfaces such as oxidized Si wafers. This is despite large dielectric constants of the tested oxides, which should suppress scattering by charged impurities.[1-4] Our observations indicate that several mechanisms contribute to charge carrier scattering in graphene and the dominant one may change for a different substrate. Nonetheless, we argue that the crucial role in achieving ultra-high electronic quality is the self-cleansing process previously reported for graphene on hBN[11] and now observed for graphene on the disulphides. In this process, van der Waals forces squeeze contamination adsorbed at contacting surfaces into sizeable pockets, leaving the rest of the interface atomically clean.[11] We expect this self-cleansing to occur for all layered TMD.[9, 13] No self-cleansing is observed for cleaved oxide substrates where contamination (including monolayers of adsorbed water[14-16]) remains distributed over the entire graphene interface.

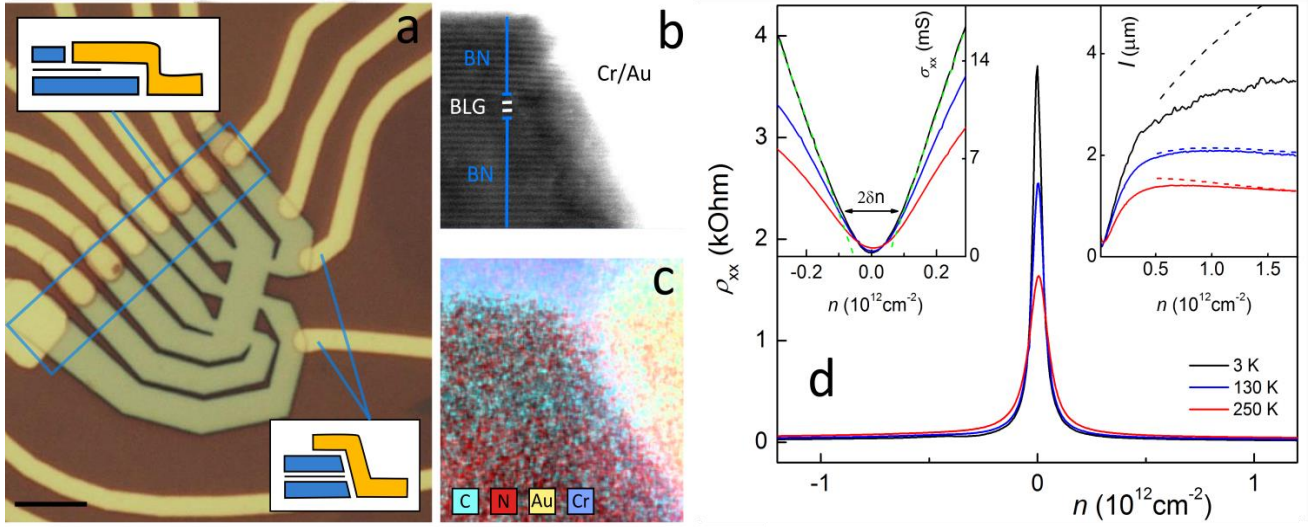

**Figure 1.** Quality of hBN/graphene/hBN heterostructures fabricated by dry peel transfer. (a) Optical micrograph of a Hall bar device with two different types of contacts: overlapping (illustrated by the top inset) and edge (bottom). The scale bar is 5 μm. (b,c) Cross-sectional TEM image of an edge contact to an encapsulated bilayer graphene (BLG) and its HAADF elemental mapping. The images are obtained using thin slices of the contact areas, which were prepared by a focused ion beam.[11] The scale is given by the interlayer distance of 3.4 Å. (d) Resistivity $\rho_{xx}$, conductivity $\sigma_{xx}$ (left inset) and mean free path $l$ (right inset) as a function of $n$ at different $T$ for the device in (a). The green dashed line in the left inset corresponds to the $1/n$ dependence and illustrates the inhomogeneity $\delta n$. The black dashed line in the right inset shows $l$ expected if no scattering occurs at device boundaries. Acoustic phonon scattering leads to shorter $l$ at elevated $T$ as shown by the red and blue dashed curves. The theory curves were calculated following refs. 7,12.

To set up a standard of electronic quality for graphene on a substrate, we start with encapsulated hBN/graphene/hBN devices. Their fabrication is described in refs. 5-12 and in Supporting Information.[17] Briefly, graphene and thin hBN crystals required for making such heterostructures were mechanically cleaved onto a film consisting of two polymer layers (PMGI and PMMA) dissolving in different solvents. We lifted the top polymer together with the chosen crystals off the wafer by dissolving the bottom layer. The resulting flake is placed onto a circular holder and loaded face down into a micromanipulation setup where it can be precisely aligned with another 2D crystal prepared on a separate wafer, which later serves as a base substrate for the final device. Unlike in the previous reports,[5-11] we no longer dissolve the PMMA carrier film but peel it off mechanically.[12] Mutual adhesion between graphene and hBN crystals is greater than either of them has with the polymer. After the transfer of graphene onto a selected crystal, the structure is immediately encapsulated with another hBN crystal (5-20 nm thick) using the same dry-peel transfer. This allows us to avoid any solvent touching critical surfaces. The final heterostructures are shaped into the required geometry by plasma etching. One of our hBN/graphene/hBN Hall bars is shown in Fig. 1a.

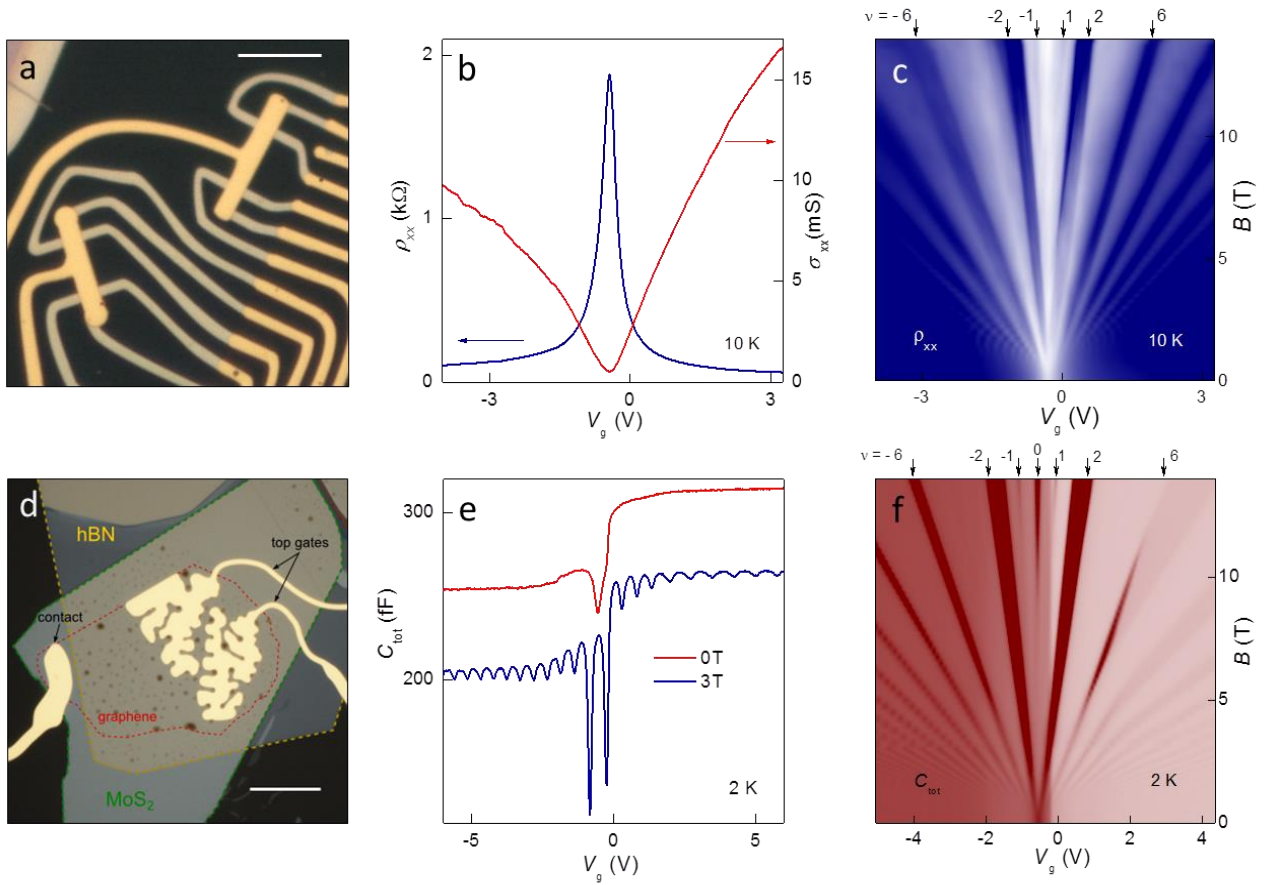

**Figure 2.** Graphene devices fabricated on a MoS$_2$ substrate. (a) Optical micrograph of a typical MoS$_2$/graphene/hBN Hall bar. The MoS$_2$/graphene heterostructure is encapsulated with a thin hBN layer that serves as a top gate dielectric. Scale bar, 10 μm. (b) Resistivity and conductivity in zero $B$ for the MoS$_2$/graphene/hBN device. (c) Its Landau fan diagram $\rho_{xx}(V_g, B)$. Scale: navy to white, 0 to ≥3 kOhm. (d) Optical image of a typical MoS$_2$/graphene/hBN/Au capacitor. The meandering shape of the top gate is to maximize the active area by avoiding contamination bubbles (dark spots). Colored dashed lines outline corresponding layers: green – is the MoS$_2$ substrate, red – is graphene and yellow is the thin encapsulating hBN, which is also used as the gate dielectric. Scale bar, 15 μm. (e) Capacitance of a MoS$_2$/graphene/hBN/Au device in zero and quantizing $B$. For clarity, the curves are offset by 50 fF. (f) Fan diagram $C_{tot}(V_g, B)$. Scale: wine to white, 0.18 to 0.3 pF. The numbers and arrows above the plot mark the filling factors, $\nu$.

Electric contacts to encapsulated graphene can be made using two different approaches. In the conventional one,[5-11] the heterostructures is designed in such a way that some areas of graphene are left not encapsulated and Cr/Au (4/80 nm) contacts could be deposited later (top inset of Fig. 1a). In the second approach,[12] the same metallization is evaporated directly onto the etched mesa that had no exposed graphene areas as schematically shown in the bottom inset of Fig. 1a. The latter method allows ohmic contacts with resistivity of ~1 kOhm/μm over a wide range of charge carrier densities $n$ and magnetic fields $B$, similar to traditional (top-evaporated) contacts.[5-11] The quality of 'edge' contacts is surprising because graphene is buried inside hBN and exposed by less than one nanometer along the edge. The edge geometry is visualized in Figs. 1b,c using transmission electron microscopy (TEM) and high-angle annular dark-field imaging (HAADF). The images show that graphene stays encapsulated within hBN up to the point where the metallization joins the graphene edge, making such electric contacts effectively one-dimensional.[12]

Large area graphene-hBN interfaces always exhibit contamination bubbles that arise from coagulation of a hydrocarbon and other residue trapped between graphene and hBN[11] (see below). Bubbles lead to significant charge inhomogeneity and, therefore, should be avoided within an active area of a device. Without the use of dry-peeling, we can usually fabricate Hall bars with a typical width of ~1 µm. In this case and for $\mu$ >100,000 cm$^2$ V$^{-1}$ s$^{-1}$, the mean free path $l$ at low temperatures ($T$) becomes limited by electron scattering at graphene edges.[7] Transport and capacitance characteristics of such devices were extensively described in literature, and we avoid repeating this information by referring to our earlier report[7, 18] and focusing below on ultra-high-quality devices obtained by dry-peel transfer.

The use of the latter approach is found particularly important because this results in less contamination, allowing bubble-free areas larger than 100 µm$^2$. Consequently, we could make homogeneous graphene-on-hBN devices up to 10 µm in size. Figure 1d shows longitudinal resistivity $\rho_{xx}$ for one of our Hall bars obtained by dry-peel transfer. In this device, the field-effect mobility $\mu_{FE} = 1/\rho_{xx}ne$ reaches $\approx$ 500,000 cm$^2$ V$^{-1}$ s$^{-1}$ at $n$ <2x10$^{11}$ cm$^{-2}$ and $T$ <20 K. This allows the mean free path $l$ of about 4 µm as shown in the inset of Fig. 1d, and $l$ is limited by the device's width. At room $T$, $\mu_{FE}$ decreases to $\approx$150,000 cm$^2$V$^{-1}$s$^{-1}$ because of phonon scattering.[12] Importantly, this electronic quality is typical rather than exceptional[12] for our large-area encapsulated devices. Note that, although $\mu$ approaches the highest values demonstrated for suspended graphene devices, the charge inhomogeneity for hBN/graphene/hBN is still rather large ($\delta n \geq 10^{10}$ cm$^{-2}$, see Fig. 1d for the definition), an order of magnitude higher than that observed in suspended graphene.[19, 20]

In search for encapsulation materials alternative to hBN, we have tried a large number of cleavable layered crystals but so far could not achieve the quality of our best hBN/graphene/hBN devices. The second best materials we found are MoS$_2$ or WS$_2$. Both disulphides exhibit high stability under ambient conditions, good chemical resistance, allow flat areas of a sub-mm size without atomic terraces and can be mechanically cleaved down to a monolayer.[9, 13, 21, 22] Because these semiconductors have a relatively narrow bandgap of ~1.5 eV and, in addition, available crystals are often doped, a gate voltage applied through the substrates is efficiently screened by accumulation and depletion surface layers.[17] This prevents the use of TMD as gate dielectrics and, more specifically, as substrates in the standard graphene geometry with a back gate.[1-8] Nonetheless, it is possible to use semiconducting crystals as substrates if the gate voltage $V_g$ is applied through a top dielectric layer. In the top-gate geometry, substrate's screening plays little role as long as there is no electric contact with graphene (see below).

Examples of our top-gated MoS$_2$/graphene/hBN devices are shown in Fig. 2. Their transport characteristics are presented in Figs. 2b,c. One can see that zero-$B$ resistivity $\rho_{xx}$ exhibits the

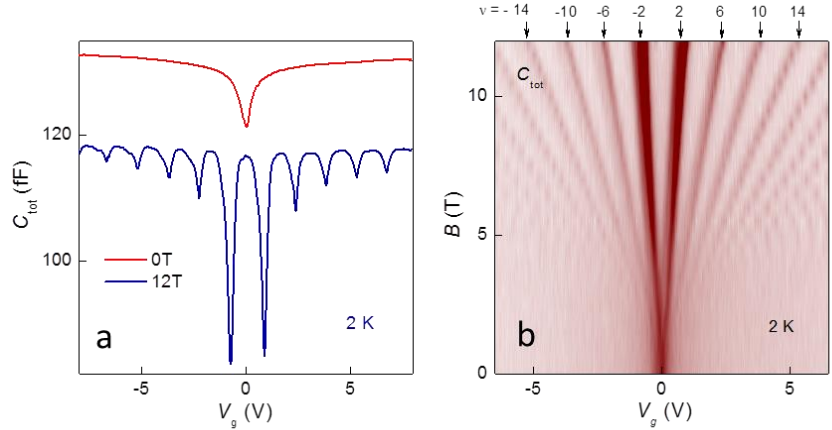

**Figure 3.** Capacitance spectroscopy of WS$_2$/graphene/hBN/Au. (a) $C_{tot}$ as a function of $V_g$ in zero and quantizing $B$. For clarity, the 12T curve is offset by 15 fF. (b) Landau fan diagram $C_{tot}(V_g, B)$. Scale: wine to white 0.105 to 0.120 pF.

standard behavior with a sharp maximum and small (~1x10$^{10}$ cm$^{-2}$) residual doping. The linear part of conductivity $\sigma_{xx} = 1/\rho_{xx}$ yields $\mu_{FE} \approx 60,000$ cm$^2$ V$^{-1}$ s$^{-1}$ (Fig. 2b), in agreement with the Hall-effect mobility measured for the same device. By applying magnetic field, we find that the quantum Hall effect (QHE) fully develops in a few T and graphene's spin-valley degeneracy is lifted at $B > 8$ T (Fig. 2c). The onset of Shubnikov - de Haas oscillations occurs at ≈0.5 T, which allows us to estimate the quantum mobility $\mu_q$ as ≈20,000 cm$^2$ V$^{-1}$ s$^{-1}$ ($\mu_q = 1/B_s$ where $B_s$ is determined as the field where additional extrema due to Landau quantization are observed[23-25]). Unlike $\mu_{FE}$ that is limited by large-angle scattering, $\mu_q$ is sensitive to small-angle scattering events that destroy coherence on quantized orbits. Therefore, it is little surprise that the two mobilities differ, and the observed factor of 3 difference agrees with the results for standard graphene-on-SiO$_2$ devices.[23, 25]

Another tool that we have employed in search for quality substrates was capacitance spectroscopy.[26, 27] The technique probes directly the density of states (DoS) and provides information about the electronic spectrum, which is difficult to extract from transport measurements.[18, 27-29] Additional advantages of using capacitance spectroscopy are that capacitor devices do not require plasma etching or multiple electric contacts, and that large area devices (>300 μm$^2$) can be made free from contamination bubbles by carefully shaping the top gate as shown in Fig. 2d. We used Andeen-Hagerling AH2550A capacitance bridge with an excitation of 5meV or lower. We measured capacitance at different excitation frequencies (0.1-10kHz) to ensure that the contribution of the spreading resistance is negligible.[17]

The total capacitance $C_{tot}$ per unit area of MoS$_2$/graphene/hBN/Au (here Au represents the second electrode of a capacitor) devices can be represented by the series connection of their geometrical $C_{geom}$ and quantum capacitances[18]

$$C_{\text{tot}}^{-1} = C_{\text{geom}}^{-1} + \left(\frac{8e^2\pi|\varepsilon_{\text{F}}|}{h^2 v_{\text{F}}^2} + e^2 D_{\text{s}}\right)^{-1}$$

where $C_{\text{geom}} = \varepsilon\varepsilon_0/d$, $\varepsilon$ is the dielectric constant, $d$ the thickness of the top-gate dielectric, ε_F the Fermi energy and $v_F$ the Fermi velocity. To account for charge carriers that may leak from graphene into a conducting substrate, we have introduced an additional term $D_S$ that is absent for graphene on hBN.[18, 30] Fig. 2e shows $C_{\text{tot}}$ as a function of top-gate voltage for a MoS$_2$/graphene/hBN/Au capacitor. It exhibits a large cusp at $V_g \approx$ -0.6 V ($n \approx 10^{11}$ cm$^{-2}$), which marks a minimum in the DoS at the Dirac point. This behavior is standard for graphene[18, 28] but now appears on top of a step-like increase in capacitance near zero $V_g$. We attribute the latter feature to a finite Schottky barrier between graphene and MoS$_2$. Indeed, electric charges can efficiently move between the two electrodes if the barrier resistance $R$ is smaller than $1/C_{\text{tot}} \times f$, where $f$ is the measurement frequency. In our case ($f$ =1-10 kHz and $C_{\text{tot}}$ ~0.1 pF), $R$ ~$10^9$ Ohm would already provide sufficient coupling between graphene and MoS$_2$ to result in a notable $D_S$ but would not be discernable as a parallel conduction in transport experiments. The graphene-substrate coupling allows the electric field created by the top gate to partially penetrate through graphene into the MoS$_2$ substrate. The observed charge accumulation at positive $V_g$ implies that the MoS$_2$ substrate is *n*-doped, in agreement with independent measurements of our MoS$_2$ crystals.[17]

    To assess electronic quality of MoS$_2$/graphene/hBN/Au capacitors, we apply a magnetic field. Fig. 2e shows pronounced oscillations in $C_{\text{tot}}$ which appear due to Landau quantization of graphene's DoS.[18] In agreement with the transport data in Fig. 2c, magneto-oscillations start at $B_s \approx$0.5 T, yielding $\mu_q \approx$20,000 cm$^2$ V$^{-1}$ s$^{-1}$, and the spin-valley degeneracy is lifted for $B$ >8 T (Fig. 2f). The step-like contribution to the DoS from the MoS$_2$ substrate remains unaffected by $B$ (Fig. 2e). Note that Landau levels in Figs. 2c,f exhibit an asymmetric behavior (slopes are steeper for positive $V_g$) and their positions in $B$ evolve nonlinearly with increasing $V_g$ (see changes in the slopes at positive $V_g$ in Fig. 2f). Such behavior is neither observed nor expected for graphene on dielectric substrates[18, 28] and appears due to the fact that some of the charge induced by electric field escapes from graphene into the MoS$_2$ substrate as discussed above ($D_S$ depends on $V_g$). The comparison of transport and magnetocapacitance measurements in Fig. 2 shows that both provide fairly similar information about graphene's electronic quality (also, see ref. 28). Because capacitors are quicker and easier to fabricate and examine, we tend to employ them more than Hall bars in testing various substrates, only checking our conclusions by transport measurements if necessary.

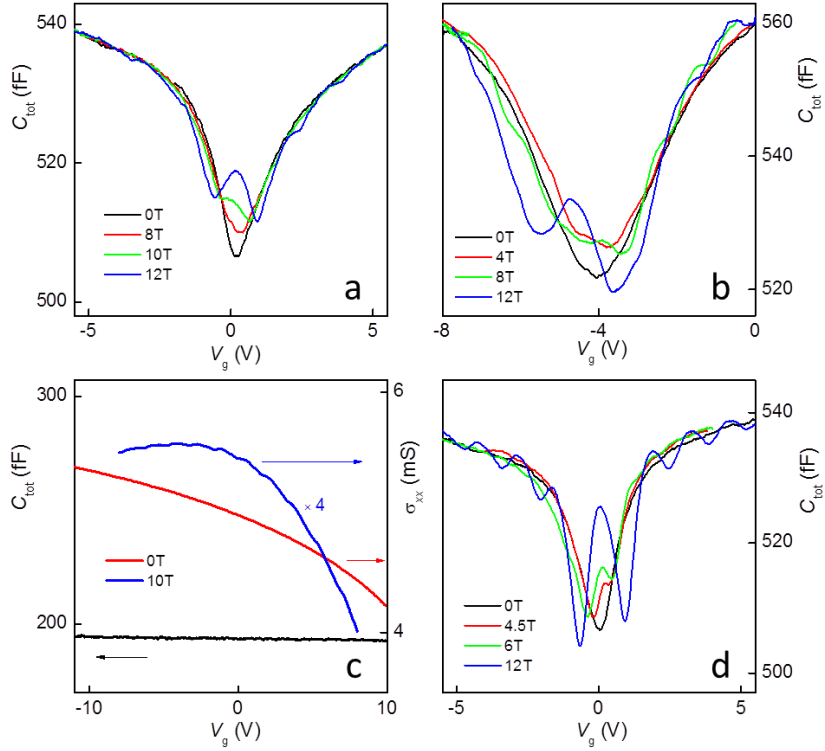

**Figure 4.** Capacitance spectroscopy of graphene on various atomically flat oxides at 2 K. (a) $C_{tot}$ behavior for a mica/graphene/hBN/Au capacitor in different $B$. The onset of magneto-oscillations occurs at $B_s \approx 10$ T. (b) Same for BSCCO/graphene/hBN/Au. (c) Capacitance (black curve) and conductivity $\sigma_{xx}$ (red and blue) as a function of $V_g$ for $V_2O_5$/graphene/hBN/Au devices. The high-$B$ curve (blue) is magnified for clarity to reveal magneto-oscillations. The charge neutrality point is shifted to large positive voltages due to heavy doping by the substrate. (d) $C_{tot}$ for graphene capacitor on atomically rough $LiNbO_3$. The onset of magneto-oscillations occurs at $B_s \approx 4$ T.

Graphene encapsulated between $WS_2$ and hBN is found to exhibit quality similar to that of $MoS_2$/graphene/hBN structures. Examples of our capacitance measurements for $WS_2$ substrates are shown in Fig. 3. In this particular device, the onset of magneto-oscillations is observed at ≈1 T, which implies $\mu_q$ ~10,000 cm$^2$ V$^{-1}$ s$^{-1}$, a factor of 2 lower than $\mu_q$ in $MoS_2$/graphene/hBN in Fig. 2. However, this is within reproducibility of our heterostructures, and another graphene-on-$WS_2$ device (Hall bar) exhibited $\mu_{FE} \approx 55,000$ cm$^2$ V$^{-1}$ s$^{-1}$, similar to mobilities observed for $MoS_2$/graphene/hBN. Fig. 3 also shows that the use of $WS_2$ substrates allows us to avoid the obscuring steps in $C_{tot}(V_g)$ and the asymmetry in the Landau fan diagrams, which were consistently present in the case of $MoS_2$. We explain this by the fact that our $WS_2$ crystals are insulating (undoped)[17] which increases the Schottky barrier and suppresses their electric coupling with graphene. Note that the use of substrates containing heavy elements may in principle lead to a proximity-induced spin-orbit gap in graphene.[9] Although graphene on the disulphide substrates exhibits positive magnetoresistance in small $B$[17] (instead conventional weak localization), which indicates notable spin-orbit scattering, no sign of any gap could be detected in graphene's DoS. Taking into account our typical broadening in the DoS, we estimate that the proximity induced spin-orbit gap should be <20 meV even for WS$_2$.

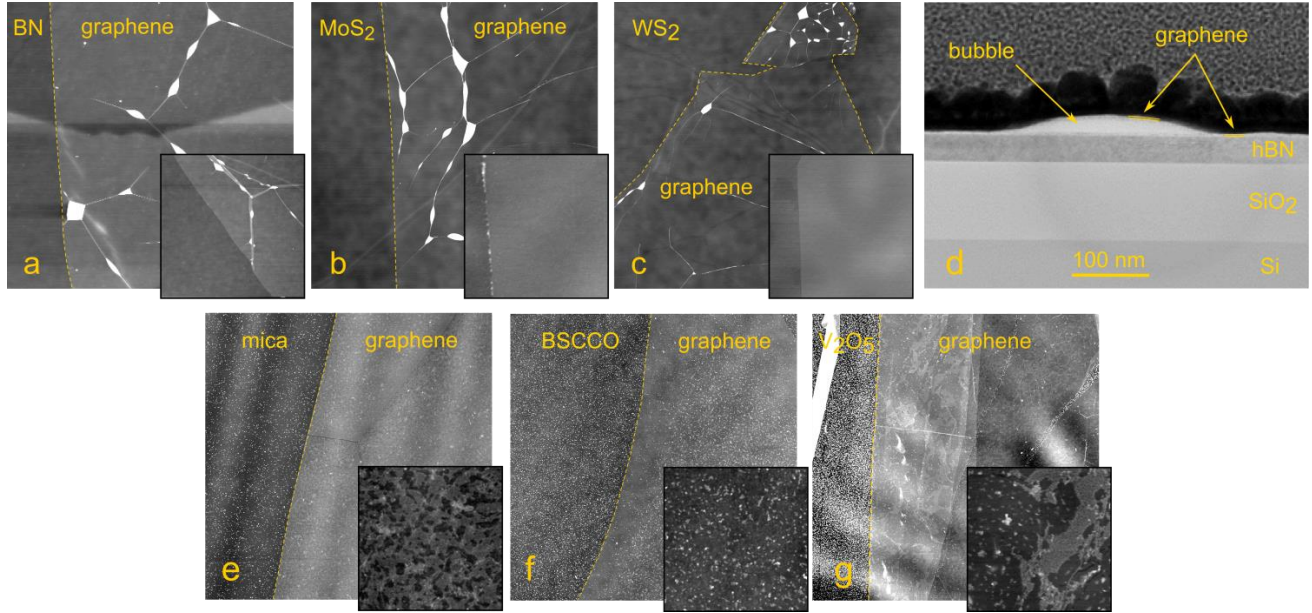

**Figure 5.** AFM topology and TEM cross section image of graphene on various substrates. (a-g) Graphene on hBN, MoS$_2$, WS$_2$, mica, BSCCO and V$_2$O$_5$. All AFM scan sizes: 15 μm × 15 μm. Scale: black to white, 5.5 nm. The yellow dashed curves highlight edges of graphene flakes. Insets: 1.5 μm × 1.5 μm; black to white, 4 nm. Due to self-cleansing for graphene on hBN, MoS$_2$ and WS$_2$, hydrocarbon contamination is aggregated into bubbles seen in (a-c) as bright spots connected by graphene wrinkles. (d) TEM cross section image of the graphene-on-hBN structure illustrating the hydrocarbon contamination bubble. The submicron-scale structures seen in the insets for (e,g) are probably due to a few monolayers of captured water.[14-16] The small bright dots in (e-g) are a residue that is due to the use of acetone to dissolve PMMA at the final stage and is not at the graphene interface.

Markedly poorer quality is found for all graphene devices encapsulated with atomically flat oxide substrates. Fig. 4 provides examples of our tests for graphene on mica, BSCCO and V$_2$O$_5$. For the mica/graphene/hBN structure, measurements in Fig. 4a reveal the onset of magneto-oscillations at ≈10 T, which yields $\mu_q$ of only ≈1,000 cm$^2$ V$^{-1}$ s$^{-1}$. Despite such strong scattering, graphene on mica is practically undoped (the DoS minimum is near zero $V_g$; $n \leq 10^{11}$ cm$^{-2}$), which is surprising and disagrees with the earlier Raman studies that inferred heavy $p$-doping for graphene on muscovite mica (≈10$^{13}$ cm$^{-2}$).[15] Similarly low $\mu$ are observed for BSCCO/graphene/hBN in both transport and capacitance measurements ($\mu_q \approx \mu_{FE} \approx 1,000$ cm$^2$ V$^{-1}$ s$^{-1}$). In this case, our devices exhibit $n$-doping of ≈10$^{12}$ cm$^{-2}$, which is apparent from the shift of the capacitance minimum to $V_g \approx -4$ V (Fig. 4b). The use of V$_2$O$_5$ as a capacitor substrate results in devices with very heavy $p$-doping (≈2×10$^{13}$ cm$^{-2}$). This is seen in Fig. 4c where the neutrality point in graphene could not be reached and its position was inferred by extrapolating the functional behavior of $1/\rho_{xx}(V_g)$ and $C_{tot}(V_g)$. Weak Shubnikov – de Hass oscillations could be observed in $B >10$ T (Fig. 4c), which allows an estimate for $\mu_q$ as $\approx 1,000$ cm$^2$ V$^{-1}$ s$^{-1}$. The values of $\mu$ found for the atomically flat oxides are up to an order of magnitude lower than those typical for graphene placed on oxidized Si wafers that have an atomically rough surface[1, 4]. To reiterate this point, Fig. 4d shows magnetocapacitance for graphene on a LiNbO$_3$ wafer that was polished but not atomically flat.[17] In this case, we find $\mu_q \approx 2,500$ cm$^2$ V$^{-1}$ s$^{-1}$, similar to graphene on SiO$_2$[4] and notably higher than the values obtained using atomically flat oxides.

We attribute the huge difference between different atomically flat substrates to the self-cleansing process that occurs at graphene interfaces with lipophilic hBN, MoS$_2$ and WS$_2$, and does not occur for oxides that are hydrophilic. In the former case, large areas of graphene become contamination free[11] ensuring little electron scattering. To support this argument, we have carried out atomic force microscopy (AFM) of graphene on all the reported substrates (Fig. 5). The large contamination bubbles that appear due to segregation of hydrocarbons, and were previously reported for graphene on hBN,[11] are also observed for graphene on MoS$_2$ and WS$_2$. They are seen in Figs. 5a-c as the bright white spots separated by extended flat regions. The latter are also shown at higher magnification in the insets. Figure 5d shows the TEM cross section image of one of these contamination bubbles. The flat regions exhibit a root-mean-square surface roughness of <0.1 nm for all the three shown cases, being limited by accuracy of our AFM. The flatness is the same as observed for cleaved graphite, hBN, mica, disulphides prior to the deposition of graphene. On the other hand, AFM images of graphene on hydrophilic oxide surfaces are radically different. They exhibit no large bubbles and surface roughness of up to a few nm (Figs. 5e-g). In the case of mica (Fig. 5e), our observations are consistent with those in refs. 14-16, which reported 1 to a few monolayers of water trapped between graphene and mica (although the samples were thoroughly annealed in the process, the final assembly was done in the ambient clean room atmosphere with 35% humidity). Dipoles within intercalating water and/or roughness and strain induced by water terraces (separated by a few tens of nm) lead to additional electron scattering and can be responsible for the observed low quality of graphene-mica heterostructures.

We believe that a similar scenario takes place for graphene on strongly hydrophilic V$_2$O$_5$,[31] although the heavy $p$-doping may indicate additional scattering by uncompensated interface charges. The same applies to atomically flat BSCCO, although in addition to hydrophilicity it exhibits some structural degradation under ambient conditions. Let us also mention that we were unable to use the described dry-peel transfer for the studied oxides because graphene adheres to their surfaces weaker than to the PMMA membrane presumably due to intercalating water. Therefore, PMMA had to be dissolved in acetone as for the standard graphene-on-SiO$_2$ and early graphene-on-hBN devices.

In conclusion, using transport and magnetocapacitance measurements, we have assessed electronic quality of single-layer graphene devices fabricated on various atomically flat substrates. Although the mobilities achieved so far for graphene encapsulated with layered disulphides are lower than those for the state-of-the-art hBN/graphene/hBN heterostructures, they are comparable to those demonstrated in early graphene-on-hBN devices. The lower quality may be due to vacancies and impurities present at or near $MoS_2$ and $WS_2$ surfaces. Nonetheless, we would expect higher quality if the disulphide devices could have been annealed at temperatures of ~300 °C, similar to graphene on hBN. Unfortunately, we find that $MoS_2$/graphene/hBN and $WS_2$/graphene/hBN devices experience a sharp decrease in mobility and homogeneity after annealing above 150 °C. The use of atomically flat oxides results in consistently low quality of graphene. The observed differences between hydrophilic and lipophilic substrates are attributed to their different affinities to graphene, which results in self-cleansing for lipophilic interfaces and its absence for hydrophilic ones. This suggests that other layered dichalcogenides can also serve as high quality substrates for graphene and rules out atomically flat oxides.

ASSOCIATED CONTENT

**Supporting Information.** The section contains the details of the dry peel transfer technique, brief description of the experimental set-ups, information about 2D substrate materials used in this study, table summarizing fabrication techniques and sample electrical quality, additional magnetoresistance data obtained in small magnetic fields as well as additional data on CV spectroscopy of bare $MoS_2$ and $WS_2$ crystals. This material is available free of charge via the Internet at http://pubs.acs.org.

ACKNOLEDGMENTS

This work was supported by the European Research Council, Graphene Flagship, Engineering and Physical Sciences Research Council (UK), the Royal Society, US Office of Naval Research, US Air Force Office of Scientific Research, US Army Research Office.

**Corresponding Author**

*E-mail: andrey.kretinin@manchester.ac.uk

# Supporting Information

## Electronic Properties of Graphene Encapsulated with Different Two-Dimensional Atomic Crystals

A. V. Kretinin[1], Y. Cao[1], J. S. Tu[1], G. L. Yu[2], R. Jalil[1], K. S. Novoselov[2], S. J. Haigh[3], A. Gholinia[3], A. Mishchenko[2], M. Lozada[2], T. Georgiou[2], C. R. Woods[2], F. Withers[1], P. Blake[1], G. Eda[4], A. Wirsig[5], C. Hucho[5], K. Watanabe[6], T. Taniguchi[6], A. K. Geim[1,2] and R. V. Gorbachev[1]

[1]Centre for Mesoscience and Nanotechnology, University of Manchester, Manchester M13 9PL, UK

[2]School of Physics and Astronomy, University of Manchester, Oxford Road, Manchester, M13 9PL, UK

[3]School of Materials, University of Manchester, Oxford Road, Manchester, M13 9PL, UK

[4]Graphene Research Centre, National University of Singapore, 6 Science Drive 2, Singapore 117546

[5]Paul Drude Institut für Festkörperelektronik, Hausvogteiplatz 5-7, 10117 Berlin, Germany

[6]National Institute for Materials Science, 1-1 Namiki, Tsukuba, 305-0044 Japan

## 1. Dry-peel transfer

The essential steps of the transfer procedure are illustrated in Fig. S1. The assembly starts with the standard mechanical exfoliation of graphene or a thin crystal of another layered material (e.g., hBN or mica) on top of a polymer stack consisting of PMGI (MicroChem SF6) and PMMA (AllResist 672.08) layers (250 nm/1000 nm). It was noticed that prior preheating of the polymer stack on a hotplate to ~ 100 °C would increase the area of the exfoliated flakes, Fig S2. The bottom PMGI layer is then selectively etched with a water-based solvent (MICROPOSIT® MF-319). The solvent is positioned around the polymer stack and does not come in contact with the transferred crystal leaving its surface dry and clean (Fig. S1). The top surface of the hydrophobic PMMA film also remains dry. The floating membrane is then picked up on a metal ring and allowed to dry up. The ring is loaded into a micromanipulation setup and aligned with a second 2D crystal chosen for the assembly.

Subsequent steps depend on a chosen wafer where the second crystal is prepared and its size. If an oxidized Si wafer is used as a substrate, the second crystal usually exhibits weak adhesion and, therefore, can be picked up by the first crystal attached to the PMMA membrane. In this case, the heterostructure can be assembled top to bottom, similar to the method described in Ref. [1]. The fully assembled stack is then deposited onto a final substrate by dissolving the carrying PMMA membrane in acetone. Another scenario takes place if the second 2D crystal has strong adhesion to the substrate. Then, the first crystal is released by the PMMA carrier film and deposited on top of the second crystal (Fig. S1). Thus, the heterostructure can be built from bottom to top as shown in Fig. S1.

In contrast to Ref. [1] we find that a choice of polymers and substrates is not important as long as no liquid processing is involved before graphene is sealed between 2D crystals. Also, in our experience, the heating during the transfer procedure is not essential, although it may help to achieve larger separation between contamination bubbles and provide better adhesion. The optimal temperature range for the heated transfer procedure was found to be 60 – 70 °C. At higher temperatures the heat convection from the substrate softens the PMMA membrane, which makes it less stable during the flake alignment. The success rate of the flake transfer onto substrates described in the main text is close to 100%.

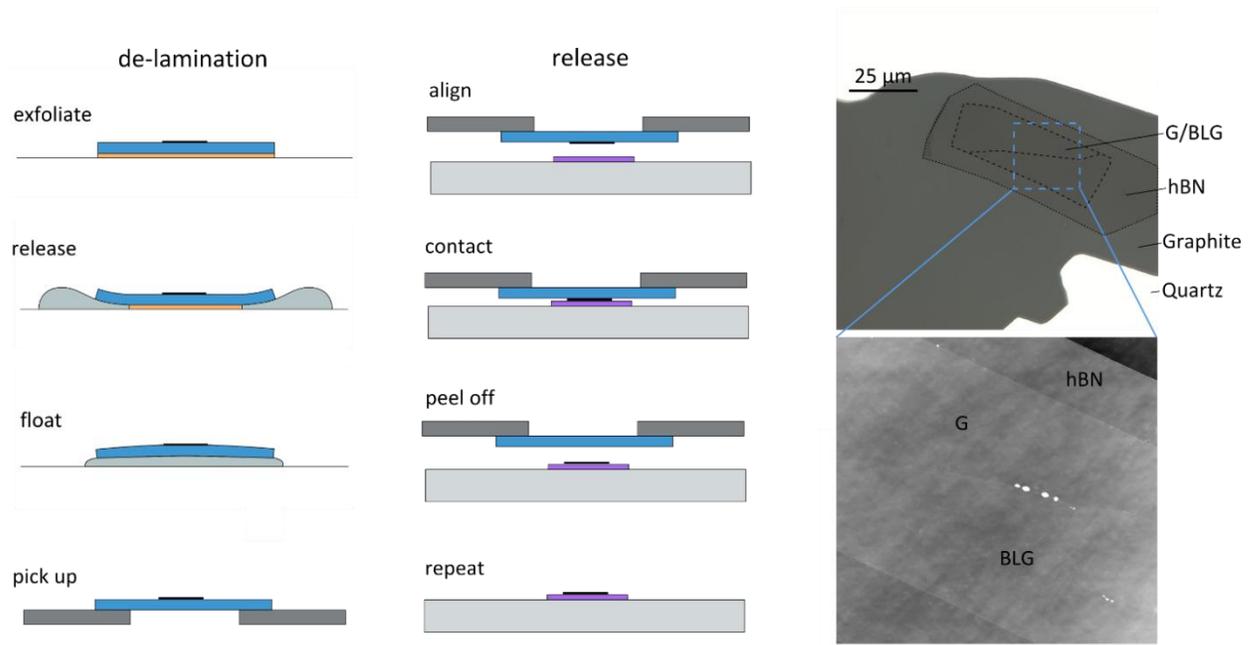

**Figure S1.** Dry-peel transfer. Left - Different steps of the transfer procedure. Right - Optical micrograph of non-encapsulated graphene/hBN/graphite heterostructure (top); Zoom-in by using AFM in the tapping mode (bottom).

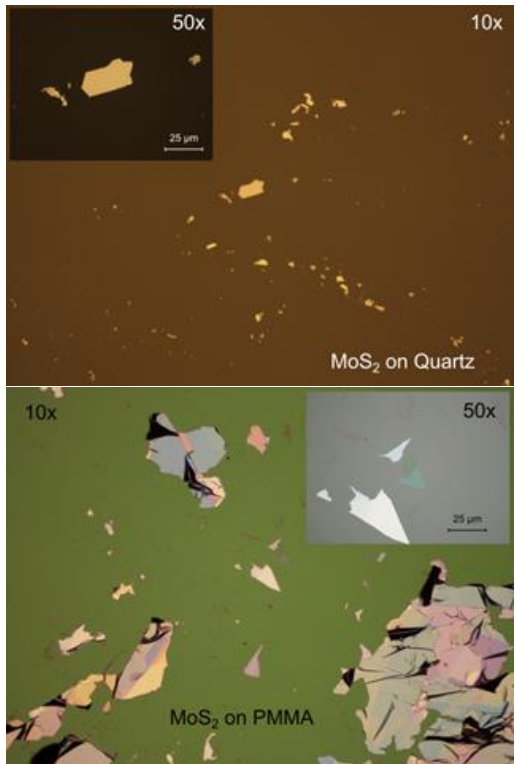

**Figure S2.** Examples of mechanically exfoliated $MoS_2$ flakes on quartz (top) and PMMA (bottom) substrates. The yield of exfoliation is very similar to that of graphene. $WS_2$ demonstrates very similar behavior.

The final devices were shaped by $CHF_3/O_2$ RF plasma etching (Oxford PlasmaLab) through a metal mask patterned by e-beam lithography.

## 2. Measurement setup

All the measurements described in the main text were carried out in a variable temperature insert inside a liquid $^4$He cryostat fitted with a superconducting magnet. For transport measurements, we employed the standard lock-in technique with constant currents of ~100 nA and at low frequencies (6-30 Hz). Capacitance spectroscopy was performed using capacitance bridge AH2550A (Andeen-Hagerling) at 0.1-10 kHz and an excitation voltage of ~5 mV. The sample was wired with coaxial cables for better control of the parasitic stray capacitance. All measurements were performed as recommended by the AH2550A capacitance bridge user manual. In order to avoid parasitic capacitances, our capacitor devices were fabricated on top of quartz wafers. The range of gate voltages, $V_g$, applied to a particular device was dictated by dielectric strength of the hBN layer limited by typically 0.5 V/nm [2,3].

### 3. Substrate materials

As substrates we used $MoS_2$ in the form of natural molybdenite (TX Materials), synthetic $WS_2$ [4], quality muscovite mica (SPI Supplies), BSCCO and layered orthorhombic $V_2O_5$ grown by the floating zone method [5]. Also, polished z-cut lithium niobate ($LiNbO_3$) wafers were used as non-atomically flat reference substrates (Roditi Ltd). The exfoliation technique used for all layered materials is same as for graphene and hBN and it was carried out in air inside a humidity controlled clean room (RH = 35% at 20 °C). Normally, exfoliation onto a square inch substrate would produce a desired flake.

### 4. Summary of fabrication techniques and sample electrical quality

| Structure | Transfer technique used | Effect of Annealing* | Carrier mobility at $T$<10K ($cm^2 V^{-1} s^{-1}$) | |
|---|---|---|---|---|
| | | | $\mu_{FE}$ | $\mu_q$ |
| unencapsulated hBN/Gr | Membrane dissolved | Significant Improvement | before annealing: 30,000 - 40,000 | -- |
| | | | after annealing: up to 100,000 | -- |
| hBN/Gr/hBN | Membrane dissolved | Significant Improvement | before annealing: 30,000 - 40,000 | -- |
| | | | after annealing: up to 150,000 | 50,000 |
| hBN/Gr/hBN | Dry-peel | Insignificant | before annealing: 450,000 - 480,000 | ** |
| | | | after annealing: up to 500,000 | ** |
| $MoS_2$/Gr/hBN | Dry-peel | Deteriorating | 60,000 | ~ 20,000 |
| $WS_2$/Gr/hBN | Dry-peel | Deteriorating | 55,000*** | ~ 10,000 |
| Mica/Gr/hBN | Membrane dissolved | Deteriorating | -- | ~ 1,000 |
| BSCCO/Gr/hBN | Membrane dissolved | Deteriorating | -- | ~ 1,000 |
| $V_2O_5$/Gr/hBN | Membrane dissolved | None | -- | ~ 1,000 |

**Table S1.**

* The annealing was performed at constant flow (0.5 l/min) of Ar/H2 gas mixture and consisted of the following steps: 100 °C for 30 min, 200 °C for 30 min and 300 °C for 3 hours. The temperature change rate between the steps was normally set to 5 °C/min.

** The SdH oscillation were already well developed at $B_c$ = 250 mT

*** See Fig S2

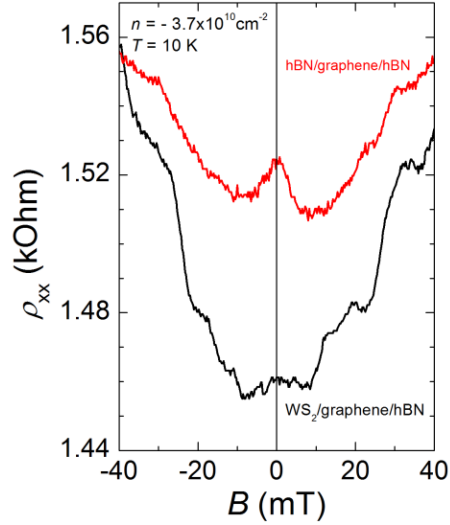

**Figure S3.** Comparison of the small-field magnetoresistance obtained from the graphene-on-hBN (red curve) and graphene-on-WS$_2$ (black curve) devices measured at the same carrier concentration.

## 5. Additional magnetoresistance data in small *B*.

Figure S3 shows the effect of different substrate on the magnetoresistance. The negative magnetoresistance due to the weak localization is typical for the graphene devices and it has been studied before.[7-9] We also reproduced this result is for the graphene-on-hBN device (red curve in Fig. S3). However, for the graphene-on-WS2 device the weak localization is suppressed (black curve in Fig S3) causing positive magneto resistance, which is attributed to the presence of the weak spin-orbit interaction [10, 11] induced by the proximity to WS$_2$.

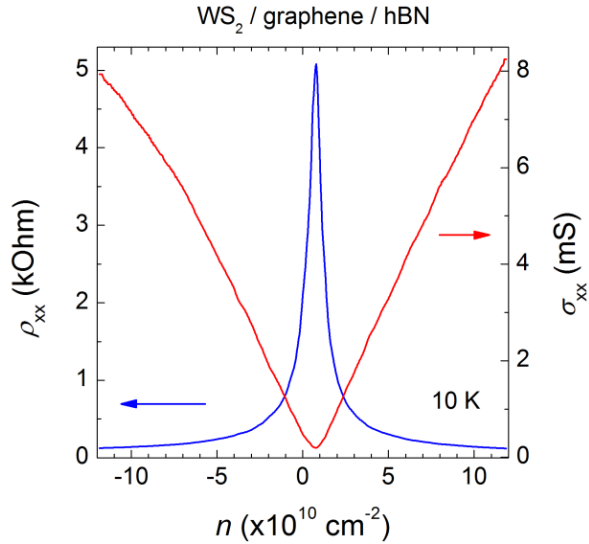

**Figure S2.** Resistivity and conductivity in zero *B* for the graphene-on-WS$_2$ Hall bar device.

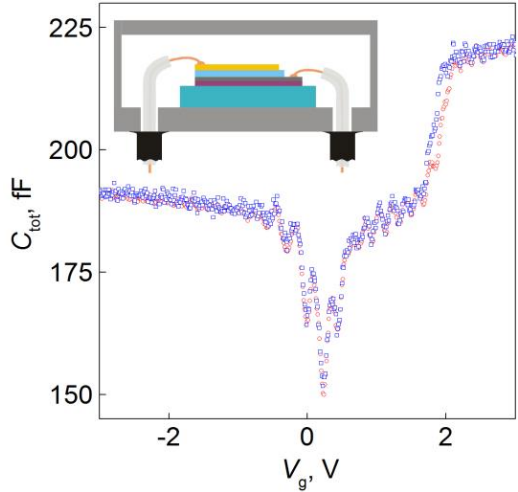

**Figure S4.** Capacitance as a function of $V_g$ for annealed MoS$_2$-based capacitor measured at different excitation frequencies. Blue dots - 0.3kHz, red dots – 3kHz. $T$=2K, $B$=1T. Annealing dopes MoS$_2$ n-type, which explains the shift of the step-like feature to positive $V_g$. Inset: schematic representation of the measurement set-up. The sample is placed inside shielded chamber and connected to the capacitance bridge by two coaxial cables (with the shields of the coaxial cables connected to the chamber).

## 6. Capacitance measurements

We used Andeen-Hagerling AH2550A capacitance bridge with an excitation of 5meV or lower. We measured capacitance at different excitation frequencies (0.1-10kHz) to ensure that the contribution of the spreading resistance is negligible, Fig. S4. The samples (prepared on quartz substrate to minimise parasitic capacitance) were placed inside a shielded chamber, and connected by coaxial cables directly to the capacitance bridge.

## 7. Capacitance spectroscopy of bare MoS$_2$ and WS$_2$ crystals

To understand the observed differences in behaviour for graphene on MoS$_2$ and on WS$_2$, we have fabricated capacitor devices similar to those described in the main text but no graphene layer was placed in between the substrates and the hBN dielectric. Figure S4a shows spectroscopy measurements for such a MoS$_2$ device at different temperatures. The observed step-like curves are typical for an *n*-type metal-insulator-semiconductor device [12]. The curves exhibit three distinct regions. The first one is the temperature independent accumulation regime at positive $V_g$, where the accumulation layer changes little so that the total capacitance $C_{tot}$ is limited by the geometrical capacitance. The second region at -1 V < $V_g$ < 0 V is an abrupt decrease in $C_{tot}$ due to depletion of the near surface layer in MoS$_2$. This is followed by the inversion region ($V_g$ < - 1 V). Here the total capacitance saturates at high negative voltages due to a finite thickness of the inversion layer, which in turn is determined by a temperature-dependent carrier concentration. Note that at low temperatures ($T$ < 180 K) the inversion changes into the "deep inversion" [13], which results in a further decrease in $C_{tot}$ and is caused by the deficit of minority carriers. The minimal value of $C_{tot}$ in this regime (≈50 fF) is limited by parasitic capacitances. At elevated temperatures ($T$ > 180 K) we also observe a minor dip at $V_g$ ~ -0.7 V, which can be associated with charging a donor impurity band [14].

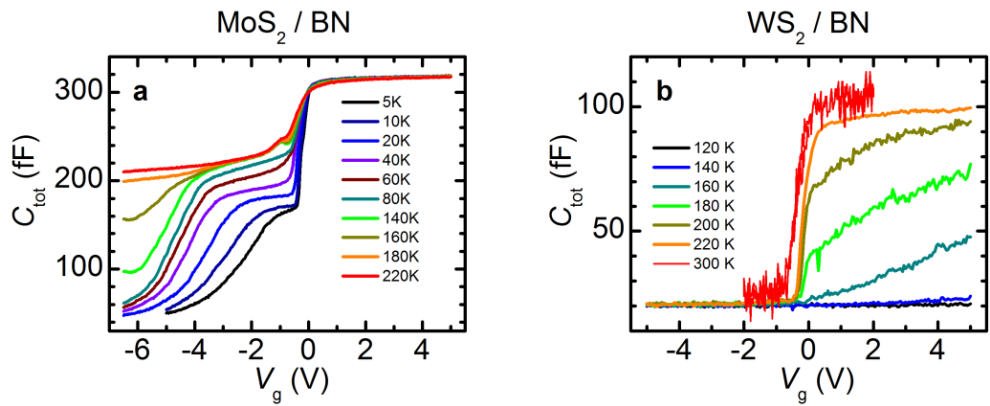

**Figure S5.** Capacitance spectroscopy of MoS$_2$ and WS$_2$. (a) Capacitance $C_{tot}$ as a function of bias $V_g$ for a typical MoS$_2$-based capacitor device (without graphene on top). (b) Capacitance measurements for a similar WS$_2$-based device.

In the case of analogous WS$_2$-based capacitors, we were unable to reach the accumulation regime (see Fig. S4b). Over a wide range of temperatures, our capacitance curves correspond to the deep inversion regime and, only at $T$ > 200 K, we could reach the normal inversion. This behaviour confirms that at low temperature our WS$_2$ crystals are good insulators with no mobile charge carriers.

**Supporting references**